# Anode initiated impulse breakdown in water: the dependence on pulse rise time for nanosecond and sub-nanosecond pulses and initiation mechanism based on electrostriction

Yohan Seepersad, Alexander Fridman, Danil Dobrynin

A.J. Drexel Plasma Institute, Camden, NJ, 08103

Drexel University, Philadelphia, PA, 19104

## Abstract

The effect of the voltage rise time on nanosecond and sub-nanosecond impulse breakdown of distilled water is studied. The dependence of anode initiated streamer inception on this parameter is shown to be more intricate than previously reported, particularly as it relates to mechanisms directly in the liquid phase. Dynamics of the emission phase for sub-nanosecond pulses with 600ps rise time are presented to enable comparison with previous work on nanosecond initiation features. Schlieren imaging is also used to show the development of optical density perturbations and rarefactions as a result of electrostriction in the liquid which were previously found for nanosecond pulses as well. The mechanism of nanopore generation in the liquid due to fast impulses proposed by Shneider, Pekker and Fridman is used to explain the results.

## 1  Introduction

The last few decades saw the applications of plasma in liquids extend well beyond pulsed-power and high-voltage insulation systems [1] (as was the main focus in earlier years) into areas of water treatment and purification [2-4], biological systems and medicine [5, 6], and nanomaterial synthesis [7, 8], among others [9]. Establishing a comprehensive model to describe in general the impulse breakdown of liquids has been very challenging however (and perhaps unachievable according to Balygin [10]), considering the strong dependence of the phenomenon on the experimental conditions [11]. Extensive reviews on the physics and mechanisms of breakdown in liquids have been performed [8, 11-17] and the results of these underscore the point that processes involved are intricately dependent on the experiment conditions. The situation is easily highlighted by comparing two very similar experimental setups which yielded different results. In one case, the work of Ceccato et al. [18] employed +40kV pulses to a pin-to-plane electrode configuration submerged in water with 20ns rise time ($E_{tip} \sim 8 \text{ MVcm}^{-1}$), wherein the results showed macroscopic bubble nucleation prior to plasma inception coupled with an associated finite, statistical delay $\sim 5 - 6\mu s$ to streamer inception. On the other hand, work of Seepersad et al. [19] utilized +24kV pulses with 4ns rise time ($E_{tip} \sim 1 \text{MVcm}^{-1}$) and revealed plasma initiation without bubble nucleation (directly in liquid) with practically zero lag to inception. While most of the experimental details were similar (liquid, gap, hydrostatic pressure, electrode material, radius of curvature), the initiation dynamics of the ensuing processes were fundamentally different.

Plasma generation in liquids *directly* has been known for a long time, though much more emphasis has been given to generalizing the breakdown thresholds in relation to the *pulse width* rather than the *rise time* (front duration) [10, 15, 17, 20, 21]. The general statement found in the literature is that a decrease in pulse width leads to an increase in streamer inception voltage and transition to direct liquid ionization processes, though this is incomplete without considering the rise time.

While this paper focuses on streamer inception in water (polar), work on direct ionization processes in non-polar liquids can be found in literature. Discharges in hydrocarbon oils were investigated in [20-22], whereby electron impact ionization (avalanche) was proposed as the primary mechanism driving discharge for both polarity tips. Direct ionization was also observed in [22] for negative pulses (cathode initiated), while bubble related ionization was prevalent for anode initiated discharge. The effect of rise time was investigated in these works as well ( see [22]), though the minimum reported value was 70ns, slower than the pulses considered in the current study. The effect of pulse rise time was also studied by Gournay and Lesaint [23] for anode initiated discharge in cyclohexane and pentane for pulses with rise time down to 10ns. These authors found that inception was preceded by the formation of dark features near the electrode tip, and a general slow increase in initiation voltage with rise time of applied pulse. At least in the range of $10 - 200$ns, the effect of pulse rise time was found to vary slowly and linearly on the initiation voltages, attributed to space charge effects. This will be discussed further in the context of the analysis of results to follow later.

In this current paper, we explicitly investigate the effect of rise time through comparing results from application of a delta shaped pules with 600ps rise time and 3ns duration to relatively longer pulses with 2 and 4ns rise times and $10 - 12$ns duration for discharge in distilled, deionized water. Direct and transmission imaging show the absence of any pre-initiation macroscopic bubbles ($\sim 1\mu m^2$), and as such only mechanisms related to direct ionization in the liquid phase will be considered to offer explanation of the results. The high permittivity of water ($\varepsilon \sim 80$) gives it unique properties and the conclusions made in from this work may not be generalizable to other non-polar liquids (transformer oil, mineral oils, hydrocarbons, liquefied gases, etc.). Only anode initiated discharge is investigated for two reasons: to eliminate the effect of electron emission by the needle tip which is a contributing initiation mechanisms in high fields at sharp tips; anode initiated streamers are often more luminous and larger [19], facilitating optical recording of plasma formation with the available apparatus.

## 2 Review of the electrostriction aided mechanism of plasma initiation in water without bubbles

The challenging aspect of generating plasma in water is the lack of room available to electrons to gain energy from acceleration by the electric field [3]. High scattering rates, unknown cross-sections, and rapid quenching [12] all contribute to the high theoretical breakdown potentials of liquids predicted by the Townsend mechanism. Extrapolation of Paschen's curve gives $\sim 30 MVcm^{-1}$ for direct ionization in water where typically the density is orders of magnitude higher ($n_{liquid} \sim 10^3 . n_{gas}$) [24]. Breakdown happens at electric fields closer to $\sim 1 MVcm^{-1}$ in reality, and generally there are two broad mechanisms used to describe plasma inception: initiation and propagation aided by macroscopic gas bubble formation, and, direct impact ionization in the liquid phase [11]. In the first case, bubbles provide low density regions filled with gas which effectively lower the breakdown threshold.

Most models describing ionization of the liquid phase directly still assume some underlying processes whereby *local density* of the liquid is reduced. The reference to a "bubble" (or lack thereof) refers to a macroscopic entity filled with vapor from the surrounding gas. In observance of Paschen's criteria, such features should be on the order of $\sim 20 - 30 \mu m$ [15, 25] to facilitate an equivalent gas-phase ionization process at $\vec{E} \sim 100 \text{ kVcm}^{-1}$. Proposed mechanisms such as thermally activated microvoids and crack propagation by Lewis [26], field assisted crack formation along field lines by Kupershtokh [27], electrostriction aided nanopore generation of Shneider and Pekker [28, 29], all claim localized reduction in density due to various effects without *bulk* changes in liquid density. The significant result is that these mechanisms predict ionization via a minimal initial perturbation to the bulk liquid. Electronic processes also should be considered (Zener breakdown, Grotthuss mechanism, field emission, field ionization, Auger process etc. [3, 17, 30]) for completeness, but are not addressed in this work. At least in the case of nonpolar liquids, it was suggested in [21] that electron avalanche processes occurred directly in the liquid phase for negative DC breakdown in cyclohexane and propane (lack of pressure dependence). No such observation was made for point anode experiments however, suggesting that field emission from the sharp tips should have been considered as well. In the case of anode initiated discharge (pertinent to this work), no such process was observed. Direct ionization in the case of water was suggested by Starikovskiy et al. in [31] which provided the motivation for this work.

The role of sub-microscopic cavitation in nanosecond-pulsed breakdown via the influence of electrostriction has been extensively pursued by Shneider and Pekker, and others [15, 19, 25, 28, 29, 32-37], both numerically and experimentally. A summary of these findings is presented in what follows.

Dielectric liquids stressed by non-uniform electric fields experience a bulk force represented by the Helmholtz equation, according to:

$$\vec{F} = e\, \delta n \vec{E} - \frac{\varepsilon_0}{2} E^2\, \nabla \varepsilon + \frac{\varepsilon_0}{2} \nabla \left( E^2 \frac{\partial \varepsilon}{\partial \rho} \rho \right) \qquad (2\text{-}1)$$

Here, $\varepsilon$ is the permittivity of the liquid with $\varepsilon_0$ the vacuum permittivity; $\rho$ is the density of the liquid; $\vec{E}$ is the electric field, and $e\delta n$ is the free-charge density. The first term in ((2-1) represents the force experienced by free charges throughout the bulk of the liquid; the second term is volumetric force associated with inhomogeneity of permittivity of the liquid. These first two terms are usually ignored if the liquid is assumed pure. The third term corresponds to electrostrictive forces acting in the liquid due to the inhomogeneous field, also taking into account changes in the permittivity of the liquid with density as dipole reorientation due to the field becomes significant. For deionized water (polar) it can be shown that the expression reduces to:

$$\vec{F} \approx \alpha \varepsilon_0 \varepsilon \nabla E^2 \qquad (2\text{-}2)$$

The force itself is independent of voltage polarity and a linear function of dielectric permittivity, which in itself explains the experimental results obtained in [32]. The authors in [29] go on to show that this electrostrictive force is significant enough to induce cavitation in the liquid as it can overcome the critical tension for rupture of the liquid (~30 MPa). In a region close to the electrode where the gradient of the electric field is highest, the electrostrictive force induces a negative pressure which leads to the generation of many nanoscopic ruptures in the liquid structure ~1 − 2nm. Such *nanopores* as they are referred to, are dissimilar to thermally activated microvoids as postulated by Lewis in [26] which arise naturally due to phonon collisions in the liquid lattice. From Lewis' analysis, weakening of liquid cohesion by the electric field could also play a role in nanopore generation near the electrode tip.

A requisite condition for electrostriction driven cavitation is a short rising front of the applied electric field and high liquid permittivity [28, 29]. The ponderomotive forces in the liquid are generated instantaneously in response to dipole polarization in the field, setting up negative pressure in the liquid. Hydrodynamic forces acting to compensate this negative pressure responds much more slowly due to the liquid inertia. Thus, as long as the voltage rising front is much faster than the time for fluid flow into the region of negative pressure, the unbalanced electrostriction pressure can cause liquid rupture. The estimated upper limit of voltage rise time for cavitation is 5ns based on numerical models [28]. Such nanosecond pulsed discharges are not associated with any measureable delay to streamer inception from the application of the high voltage [20, 25]. While delays (several 10's of ns) were reported for nanosecond pulsed breakdown in other work, pulse rise time was either not mentioned or longer than 5ns, and other slower mechanisms may have been at work.

It should be noted that the authors Ushakov and colleagues in [15] state that electrostriction plays no role in cavitation leading to impulse breakdown, though more thorough analysis of this effect took place in the following years. Instead, in [15] it is argued that the lack of a registered phase transition associated with nanosecond breakdown in liquids is a result of autoionization processes. A model invoked from the work of Kupershtokh in [38] was used to describe fluctuating dielectric strength throughout the liquid as a result of random, localized changes in $\varepsilon$ of the liquid due to polarization effects. This also qualitatively explained the fractal appearance of the streamer structure, however it poorly explained the dependence of discharge inception time on rise time.

Further work postulated that the electrostriction nanopores become elongated along the electric field lines [35, 37]. This work was fundamentally different from the analysis of dielectrophoretic forces acting on dielectric bubbles [39, 40], though resulted in the same conclusions: for a stationary bubble/void in an electric field, forces acting on the walls act to elongate a spherical feature into a prolate spheroid along the electric field lines. This effect is well studied for bubbles (gas filled). Rather than dielectrophoretic or vapor pressure forces contributing to bubble stretching, the unbalanced negative pressure at the poles. According to simulations, such elongation could stretch the nanopores up to ~10nm, which is sufficiently long to support electron acceleration up to ionization energies. Coulomb forces (which were ignored) acting on the inner surface of the pores will enhance the longitudinal stretching [40]. It is conceivable that within a few micrometers of the electrode surface, chains of these nanopores can link together forming longer channels. This model is fundamentally different from the mechanical expansion of cracks as proposed by Lewis [26], or the anisotropic decay of liquid along field lines by Kupershtokh et al. [27] which both offer little on the transient processes and its relationship to impulse rise time. The expansion of nanopores in the electrostriction model requires a finite time for elongation to appropriate size, but according to the most recent results [37], nanopore growth occurs rapidly ~0.1ns. This time scale corresponds to the observation of luminescent plasma on similar time scales along the rising front of the voltage pulse in [19, 25, 31].

There is no reason to believe that electrostriction may not cooperatively work with other mechanisms, simultaneously occurring and enhancing each other's effects. For the experimental conditions in this investigation however, we will argue that this mechanism offers the most consistency in interpreting the results, offering quantitative estimations that correlate well with observations. In this case, the mechanism becomes important in these very special set of circumstances: very fast rise times and liquids with high permittivity.

## 3   Experimental Setup and Methodology

A schematic of the experimental setup is shown in Figure 1. The plasma reactor is based on a pin-to-plane electrode configuration, submerged in 50 ml of deionized, distilled water (EMD Chemicals, *type II*) with a maximum conductivity of 0.1 µS cm$^{-1}$. Two different electrodes are used to study the dynamics and thresholds as it depends on the radius of curvature of the electrode: a $4 \pm 2$ μm radius of curvature iridium electrode (Microprobes for Life Sciences, IR20030.1A4), and a $25 \pm 5$ μm radius of curvature mechanically sharpened tungsten electrode. Three different power supplies are used; Table 1 summarizes the parameters of each, with comparative curves of the pulses shown in Figure 2 at maximum incident pulse amplitude.

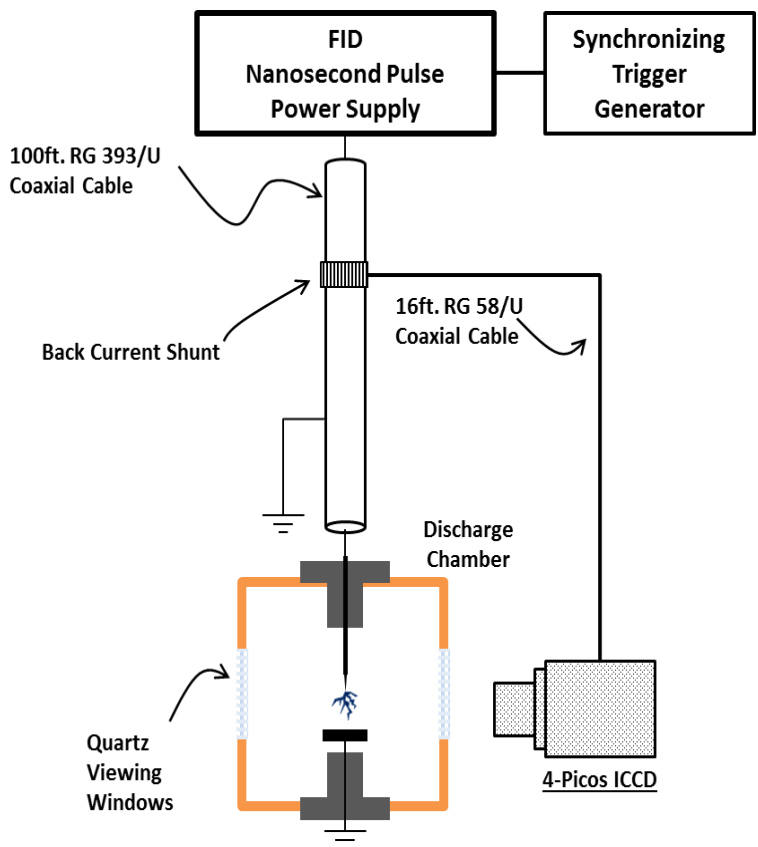

Figure 1: Schematic of basic experimental setup for monitoring plasma formation in water.

Table 1: Summary of power supply parameters used in experiments. Voltage amplitude measurements are corrected for doubling on the gap due to reflections, as well as signal attenuation. Rise time and duration are measured from the incident pulse to the electrodes.

| Power Supply Model | Rise Time $\tau_r$ (at $0-90\%$ Peak) | Duration (at 63% Peak) | Min Amplitude (On Gap) | Max Amplitude (On gap) |
|---|---|---|---|---|
| FPG 20-01 PN | $\tau_r = 0.6$ns | $2.2 - 2.3$ns | 12 kV | 15.6 kV |
| FPG 20-05 PN (Old Model) | $\tau_r = 1.6 - 2.2$ns | $10 - 12$ns | 12.5 kV | 23.5 kV |
| FPG 20-05 PN (New Model) | $\tau_r = 4 - 4.5$ns | $12 - 14$ns | 15 kV | 23.9 kV |

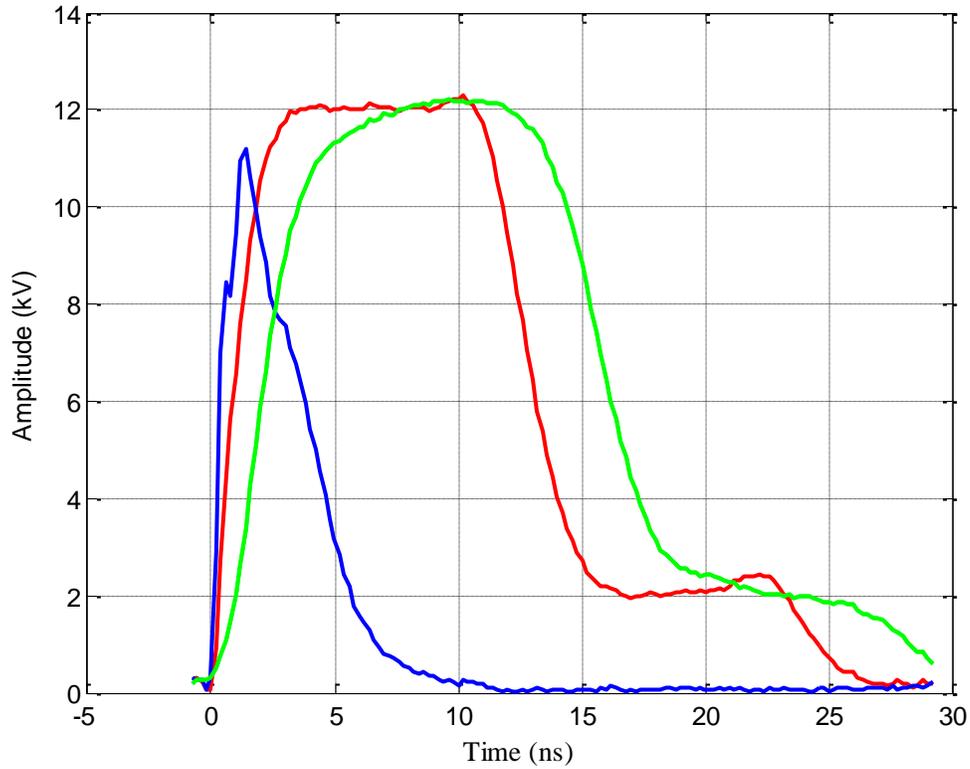

Figure 2: Incident pulses generated from power supplies listed in Table 1. Pulse amplitudes are calculated from return-current shunt measurements.

Maximum operating frequency was 10 Hz in all experiments. From visual analysis, this repetition rate was sufficiently low to allow replenishing of the liquid in the vicinity of the electrode in between pulses. We also assume that residual chemical effects from subsequent pulses were negligible on the physical mechanism since the ionic products have very short lifetimes.

Optical diagnostics were mainly employed, particularly ultra-fast ICCD (intensified charge coupled device) imaging, Schlieren imaging and shadow imaging. The camera used throughout was the 4-Picos ICCD Camera from Stanford Computer Optics. This instrument has a gating jitter better than 20 ps when triggered externally, with a minimum exposure time of 200 ps. The spectral response of the detector was $250 - 780$ nm, though absorption in the $UV$ region by water and the coupling optics should be considered. The focusing lens used allowed an image $\sim 780 \times 540 \mu m$ to be captured for direct imaging experiments. The resolving power of the entire system allowed $\sim 1 \mu m^2$ to be spatially resolved. Laser Schlieren and shadowgraphy were facilitated by a 35mW laser diode at 405nm (D6-7-405-35-M, Egismos), regulated by a temperature controlled power supply (LDC202C, Thorlabs). Even with a stabilizing power supply, the intensity of the laser illumination varied over time, and was often influenced by electromagnetic noise from the high-voltage cables and electrodes. This did not affect qualitative measurements.

A return current shunt was mounted to the middle of the high-voltage cable delivering the pulses to the electrodes; its construction and equivalent circuit can be found in [19] or [41, 42]. This signal was a direct measurement from the travelling pulse to the electrodes and was used to trigger the gating of the ICCD camera; this effectively eliminated the jitter associated with the triggering of the power supply. Better than 0.2 ns synchronization accuracy was needed, so transmission line theory was used to measure transmission line delays, as well as careful consideration of optical path delays (1ns/ft assumption for air) to the sensor. Gating delays imposed through the camera software were used to synchronize the images captured with the applied voltage at the electrodes.

The transmission line used to deliver the pulses to the electrodes was $\sim 100$ ft of RG 393/U coaxial cable. For the short, fast pulses considered in these experiments, signal attenuation becomes a significant factor when trying to estimate the actual voltage delivered to the electrodes after leaving the power supply. We estimated this peak voltage on the gap according to transmission line theory [43], by taking the voltage at the electrodes (load) as a sum of the incident voltage wave $\vec{V}^+(z,t)$ and the reflected wave $\vec{V}^-(z,t)$ (at $z = L$, $L$ is the load). Correcting the measured peak amplitudes for transmission line attenuation and using the calculations described in [44], we get a maximum peak gap voltage of ~24 kV. All voltage readings were similarly corrected in the results that follow.

## 4 Results

Optical diagnostics of the initial stage of the discharge from the FPG 20-01 PN power supply were performed since time resolved emission and pre-breakdown studied were performed previously for the other power supplies, reported in [19, 25, 31-33]. The results were used to verify two features: absence of bubbles immediately before or at the moment of discharge initiation; and, the absence of a finite delay between application of the voltage pulse and the first emission (streamer inception). Estimates of the electric field at the moment of inception are calculated, bearing in mind the width of the camera gate compared to the pulse rise time is comparable.

### 4.1 Initial stage of plasma formation for sub-nanosecond pulses

Density changes at the liquid-electrode interface were monitored using laser-backlit shadow imaging, the results of which were coupled to the ICCD images taken of only the emission without any backlight - Figure 3. The ICCD images are recorded with 200ps exposure of 50 accumulation which essentially provides an average spatial behaviour of the emission pattern. (The same technique was used in [31].) Over the first nanosecond during which emission is clearly visible, no sign of changes in the liquid density are observed larger than 1 μm. From the results, we found very faint perturbations in optical density developing after about 1.2 ns near the region of emission; these eventually develop into very thick, opaque channels after about 50 ns. Based on the ICCD imaging, these channels are radiative and can be considered similar to the plasma filaments observed in positive pulsed corona in gases. A better comparison highlighting the appearance of these dark channels is shown in Figure 4 – shadow images taken at 1 ns and 2.4 ns as indicated. On the timescale of the formation of the first radiative species in the discharge, we found that there are no large-scale (~ 1 μm) density perturbation responsible for the initiation phase of this plasma.

Note: In these pictures (Figure 3 and Figure 4), the downward directed dark filament present in each frame is a loose strand of the dielectric coating (3μm Parylene-C) on the electrode that has become frayed and detached from the electrode after multiple discharges.

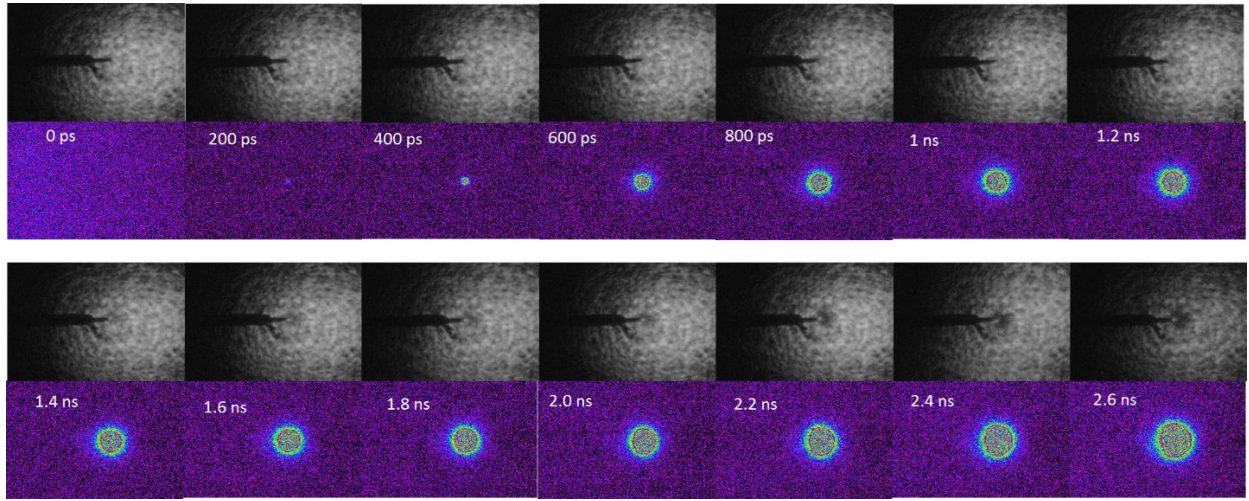

Figure 3: Shadow (single shot) and ICCD (50 accumulations) images showing the initial stage of the plasma formation. The indicated times are relative to the applied pulse in Figure 2. Image sizes are $750 \times 500$ ($\pm 10$) μm.

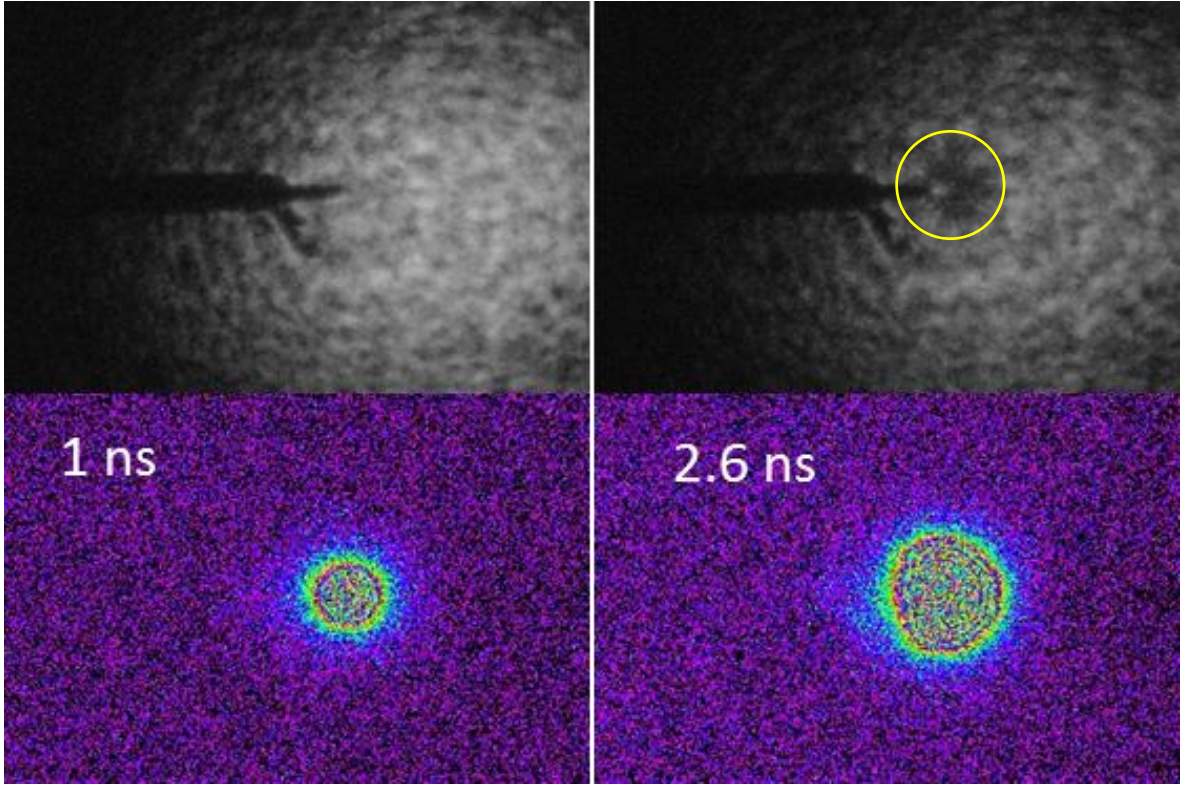

Figure 4: Shadow (top row) and ICCD (bottom row) images showing optical density changes occurring 1 and 2.6 ns after the first emission is recorded. Comparison shows the absence of any refractive index changes initially. The circle in the figure on the top-right highlights the region of refractive index change which is smaller than the emitting region.

4.2 Under-breakdown liquid-electrode dynamics for sub-nanosecond pulses

For conditions just below the breakdown initiation, the interface at the liquid-electrode tip was studied using Schlieren imaging similar to the experimental work done in [33]. A larger electrode with a radius of curvature ~ 25 µm was used to establish a larger surface area at the tip in contact with the liquid; this was necessary to provide a clearer interface for imaging the optical density perturbations due to the electric field. The larger electrode radius also resulted in a lower local electric field, since for a pin to plane geometry the electric field near the tip is related to radius by: $\vec{E} \sim U/r_0$, where $r_0$ is the radius of the tip and $U$ is the applied voltage. The applied voltage pulse was ~12 kV in these experiments, however the local electric field was and order of magnitude less than the inception voltages from the earlier experiments. The full experimental setup used in the Schlieren experiments is described in detail elsewhere [33].

The results of the Schlieren imaging experiments are shown in Figure 5. Each image was taken with a camera exposure of 1 ns, combining 20 accumulations taken at the same instant from different pulses (relative to the applied pulse) and background subtracted to improve the signal-to-noise ratio. Each image size is 300 × 200 µm, and the labels indicate the time relative to the application of the pulse with respect to the timing of Figure 2.

The results indicate that upon the application of the pulse to the electrodes, a clear change in optical density is observed in the liquid in the vicinity of the electrode tip where the local electric field is highest and most non-uniform. As reported in [33], this observation could be linked to electrostriction taking place in the liquid. The time scale over which the perturbation is recorded occurs on the order of the time within which plasma emission is first observed, though the time resolution is much coarser. Attempts to improve the time resolution of the Schlieren experiments required a more powerful illumination source which was not available at the time of the experiments. A higher powered laser source would allow an acceptable signal to noise ratio at ICCD exposures of 200 ps (camera minimum). The size of the region of perturbation recorded was on the same order of magnitude as that recorded in previous experiments with nanosecond pulses (~ 5 µm) [33], and remains in good agreement with the numerical estimates made based on the models from [28] for the geometric and electrical conditions considered in these experiments.

After 40 ns, the propagation of a compression wave away from the electrode is observed, as outlined at 50 ns and shown later at 150 ns. Such propagating waves are previously reported to be generated as a result of electrostriction [32, 36] by separate groups.

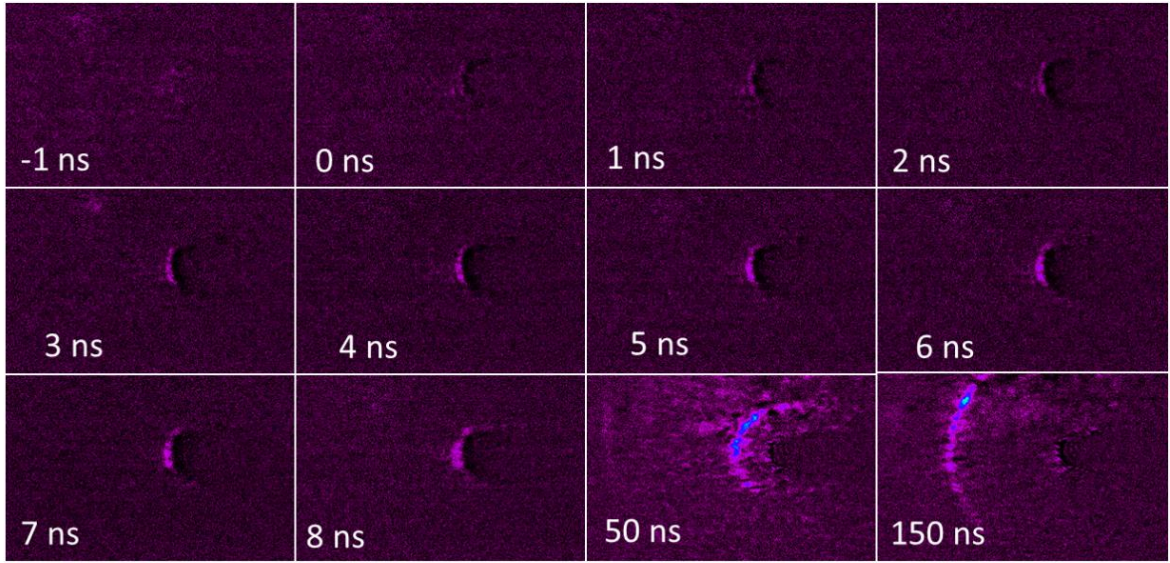

Figure 5: Schlieren imaging showing optical-density perturbations occurring at the liquid-electrode interface as a result of the applied voltage pulse. The image sizes are 300 × 200 μm and the radius of curvature of the electrode was 25 μm.

4.3 Dependence on pulse rise time

Throughout these experiments, plasma initiation (emission) occurred only on the rising edge of the applied voltage pulse. Through monitoring of the time resolved emission for each of the applied pulses from the generators in Table 1 it was verified that if the plasma did not initiate along the rising edge of the applied voltage pulse, it did not form later on. The pulse amplitude for each generator was adjusted to three criteria: maximum output voltage of the generator; voltage maximum with no plasma; voltage at threshold of initiation (probability of plasma formation ~50% per pulse). The stochastic formation processes of discharge initiation was studied by the French group of Rousseau and colleagues [13, 45] as well as in earlier work by Gournay [23] for non-polar liquids.

In these experiments, we considered that the observation of light emission corresponded to plasma initiation; this emission phase was monitored using ICCD imaging as in the earlier experiments. The inception probability was calculated as the number of events where emission was observed divided by the number of applied pulses for 100 applied pulses to the electrodes. The event of having 50% chance of initiation had a variance of about 10%. We accepted this value since changes in the amplitude of the power supplies to try to reduce the standard deviation required more sensitive control over the power supply output settings.

The applied voltage curves and corresponding discharge initiation events are shown in Figure 6. The results are summarized in table 2 following. The voltage reported is the value on the electrodes at 90% peak amplitude ($t = \tau_r$) as listed in Table 1. The electrode setup used in this experiment was a mechanically sharpened tungsten rod with a radius of curvature $20 \pm 5\mu m$ with an inter-electrode spacing of ~4mm.

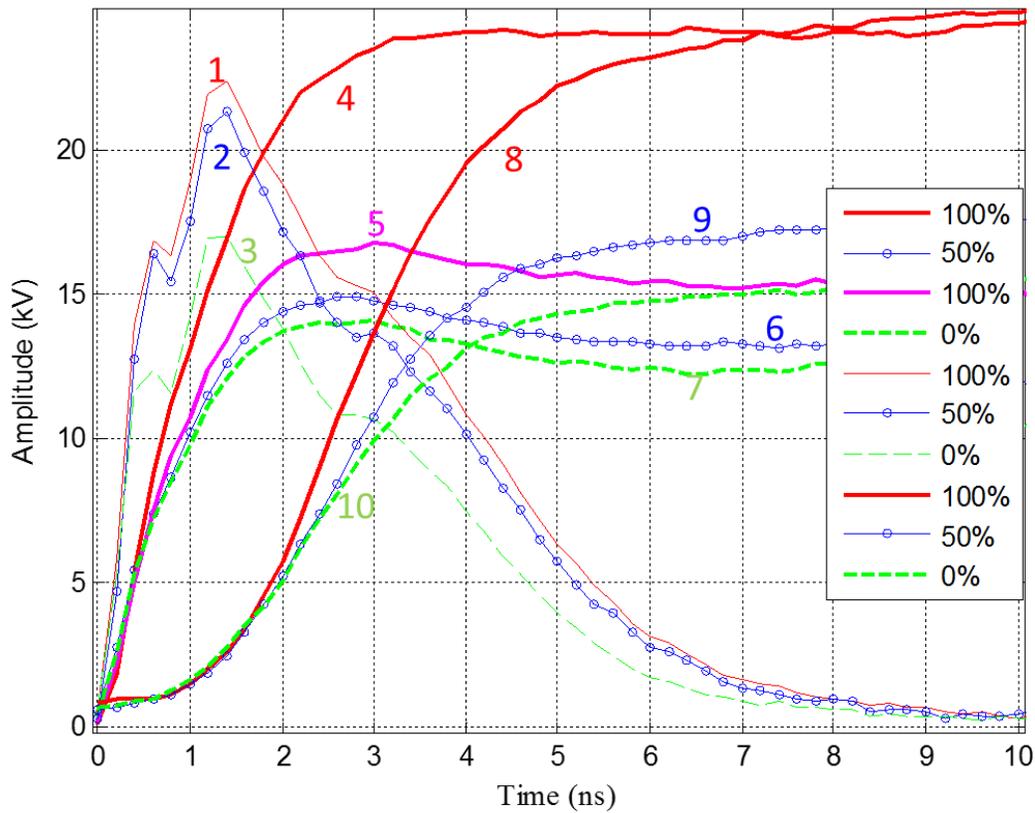

Figure 6: Applied voltage curves with probability of discharge initiation event indicated. Curves 1, 4 and 8 represent the maximum power supply output.

Table 2: Summary of initiation events for various power supply settings.

| Power Supply | Rise Time $\tau'_r$ | Amplitude at $t = \tau'_r$ | Maximum Pulse Amplitude | Probability of inception event per pulse |
|---|---|---|---|---|
| **FPG 20-01 PN** | 600ps | 16.9 kV | 22.4 kV | 100% |
|  | 600ps | 16.5 kV | 21.3 kV | 10% |
|  | 600ps | 12.3 kV | 17.0 kV | 0% |
| **FPG 20-05 PN (Old Model)** | 2.2 ns | 22 kV | 23.8 kV | 100% |
|  | 1.7ns | 13.7 kV | 15.0 kV | 50% |
|  | 1.6ns | 12.9 kV | 14.0 kV | 0% |
| **FPG 20-05 PN (New Model)** | 4.5ns | 21.1 kV | 24.8 kV | 100% |
|  | 4.2ns | 15.2 kV | 17.5 kV | 50% |
|  | 4.0ns | 13.2 kV | 15.5 kV | 0% |

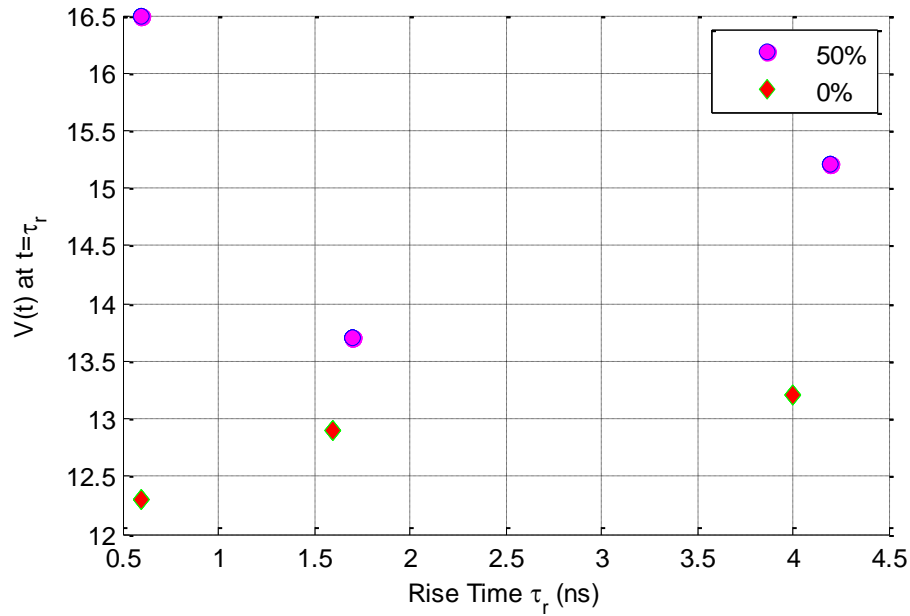

Figure 7: Summary of results form Table 2 showing pulse amplitude dependence on rise time for initiating plasma at 50% probability per pulse and no plasma.

From the results in Table 2 and Figure 7 we found that a minimum initiation voltage exists when the pulse rise time is between $2 - 4$ns. It was observed in [23] for non-polar liquids that the inception thresholds decreases slightly with pulse rise time. Although the minimum pulse rise time considered by Gournay was 10ns, the trend in the data suggests a continuous decrease in initiation voltage as rise time decreases. From our work however, we find that as the rise time in further decreased, a minimum threshold exists, after which the inception voltage increases again.

Comparing curves 6 and 10 from Figure 6, we see that both attain similar maximum voltage levels, but plasma is only initiated with the faster rising pulse. Similarly, the difference between curves 5 and 9 is also through the parameter $\tau_r$ which corresponds to $dE/dt$; curve 5 represents a pulse with a faster rate of rise of voltage than 9, however the inception probabilities are different. The change in the trend come when one compares the remaining results with curves 1, 2 and 3. Curve 2 has a higher amplitude than 5, but only initiates plasma 50% of the time, and curve 3 not at all (same amplitude as 5). Consequently, $V_{max}$ and hence ($E_{max}$ – local electric field) is found to play an equal or lesser role than $\tau_r$ which argues that $dE/dt$ plays a significant role in the breakdown mechanisms and delay times as was previously suggested in [13, 19] separately. One must remember that this dependence comes only in the unique regime of nanosecond and sub-nanosecond, anode-initiated impulse breakdown in water. Further modeling and experimental work might show whether or not this dependence applies to non-polar liquids.

## 5  Discussion

At least for the initial stage of plasma generation in water, no pre-initiation changes in density are recorded near the electrode tip as seen in Figure 3 and Figure 4, though this behavior is only consistent with previously recorded findings [31]. Clearly ionization and initiation propagation occurs directly in the liquid phase, or at least within sub-microscopic voids beneath the detection threshold. Initiation occurs immediately with the rising front, though the coarse resolution of the camera gate compared to the rising front prohibited accurately timing the moment of inception in relation to the applied pulse. The faintly opaque features forming after 1.5ns (shadow images in Figure 4) are consistently smaller than the region of emission, suggesting two things. Firstly, the propagating streamers are growing directly in the liquid phase, and not through propagating gaseous channels. Secondly, the opaque filaments forming could be gas channels beginning to develop as recombination takes place along the filaments with electrons from the head of the streamer moving along the highly conductive ionized filament and recombine with water ions. Thus, at least from the imaging of the emission phase, initiation and propagation most likely occurs directly in the liquid phase (without changes to bulk liquid properties), which then evolves into gaseous channels as dissociative recombination takes place along the ionized channel.

The noteworthy aspect however is the under-breakdown phenomena which occurs near the electrode, Figure 5. The dark region appearing in the vicinity of the electrode has been linked to electrostriction occurring in the liquid, and is consistent with previous findings and simulations [28, 32, 33]. Within 1ns (resolution of the Schlieren images), the field on the electrode is already at its maximum, and consequently, negative pressure from electrostriction is also at a maximum. Accordingly, conditions for liquid rupture are met and leads to the generation of a region saturated with nano-sized cavities which is registered as a region with lower refractive index by Schlieren imaging.

The significant aspects of the findings of this report is the delicate dependence of the inception threshold with rise time and electric field amplitude. For very fast rising pulses ($\tau_r \sim 600$ps), a higher amplitude is required to initiate plasma than at a rise time of $\sim 2 - 3$ns. Furthermore, if the rise time increases to $4 - 5$ns the amplitude needs to be even higher to initiate plasma. Beyond rise times of $5$ns, either higher amplitude pulses are required, or mechanisms related to macroscopic bubble formation begin to dominate. A minimum pulse amplitude for streamer inception exists when the rising front of the voltage is on the order of $2 - 3$ns, and this can be explained by considering the processes involved with electrostriction driven cavitation in the liquid. Two transient process take place which compete and depend on each other according to this model. Firstly, the fluid inertia must not have time to compensate the negative pressure produced by electrostriction, thus rising fronts longer than $\sim 5$ns cannot invoke cavitation (or at least to the required degree) in the liquid. In such a situation, the fluid reacts to the negative pressure and hydrodynamic flow prevents rupture. Thus, probability of nanopore appearance in the liquid increases with decrease in rise time. The second event is the growth of the nanopores, which require some finite time to become elongated in the electric field. Very fast rising pulses ($< 1$ns rise time) will facilitate the formation of nanopores which begin to elongate and link together with a finite time $\sim 1\text{ns}$). If these pores cannot grow to a sufficient dimension, a larger electric field is required to provide electrons with energy to accelerate. This approaches direct impact ionization in the limit as the rise time increases. Thus, the longer the rise time, the more the pores are allowed to grow. The minimum inception threshold will therefore occur at the moment when the pores grow to a sufficient length to facilitate ionization at the applied field.

Direct ionization mechanisms which rely only on the pulse amplitude therefore are not sufficient to explain the inception threshold dependence on pulse rise time in this regime. Field ionization processes will not be hindered by short rise times as long as the pulse amplitude is sufficient, thus it is unlikely to be a significant driving mechanism in these experiments. This comes from the fact that pulses with sub-nanosecond rise times require higher amplitude to initiate discharge than for slightly slower ($\sim 2-3\text{ns}$) rising pulses. These conditions and additional modeling work should be done with non-polar liquids to assess potentially broadening the generalization of the mechanism.

# 6 Conclusion

Streamer inception voltage in liquid water for anode initiated breakdown is more closely related to the rising front of the voltage pulse ($dV/dt$) than the peak amplitude or the pulse width. This recognition comes naturally when one considers that inception has been recorded by many groups to occur along the rising edge of the applied pulse, before maximum amplitude has been reached. The dependence is not as straightforward as one might think however, and in the range of several hundred of $\text{ps}$ to a few $\text{ns}$, a minimum inception threshold exists. The model of microscopic cavitation by electrostriction explains this subtle dependence better than previously proposed mechanisms of initiation "directly in liquids".